\documentclass{ws-jai}
\usepackage[flushleft]{threeparttable}
\newcommand{\sk}{$\widehat{SK}$~}
\newcommand{\esk}{\widehat{SK}}

\begin{document}
\catchline{}{}{}{}{} 
\markboth{Nita et al.}{EOVSA SK Correlator}

\title{EOVSA Implementation of a Spectral Kurtosis Correlator for Transient Detection
 and Classification}

\author{Gelu M. Nita$^1$, Jack Hickish$^2$, David MacMahon$^2$, Dale E. Gary$^1$}

\address{
$^1$Center for Solar-Terrestrial Research, New Jersey Institute of Technology, Newark, NJ 07102, USA, gnita@njit.edu\\
$^2$Radio Astronomy Laboratory, University of California Berkeley, Berkeley, CA, USA\\
}

\maketitle
\begin{abstract}
We describe in general terms the practical use in astronomy of a higher-order statistical quantity called Spectral Kurtosis (SK), and describe the first implementation of SK-enabled firmware in the F-engine (Fourier transform-engine) of a digital FX correlator for Expanded Owens Valley Solar Array (EOVSA).  The development of the theory for SK is summarized, leading to an expression for generalized SK that is applicable to both SK spectrometers and those not specifically designed for SK.  We also give the means for computing both the \sk estimator and thresholds for its application as a discriminator of RFI contamination.  Tests of the performance of EOVSA as an SK spectrometer are shown to agree precisely with theoretical expectations, and the methods for configuring the correlator for correct SK operation are described.
\end{abstract}


\section{Introduction}

The quality of radio astronomy scientific data can be greatly affected by radio frequency interference (RFI). With the increased use of wireless communication systems operating in frequency bands of scientific interest, the RFI environment becomes ever more hostile. At the same time, astronomers seek to observe over an ever broader part of the radio spectrum, both for increased sensitivity and to exploit spectral continuum diagnostics. These trends make it necessary to find efficient and robust methods that are able to discriminate and excise the RFI contamination, while preserving as much of the underlying useful information as possible.

There is no universal solution for the problem of RFI mitigation, and one must find or develop methods that are most suitable for a particular instrument and the RFI environment in which it operates \citep{FB2001}. Such methods include active sampling and subtraction of RFI signals, detection and excision (blanking), and spatial nulling using interferometric techniques \citep{Ardenne}.  Among these methods, RFI mitigation algorithms based on statistical discrimination of signals in both the time domain \citep{Ruf} and the frequency domain \citep{Fridman2001, FB2001} have become increasingly popular in recent years, due to the feasibility of their implementation using field-programmable gate arrays (FPGAs) or graphical programming units (GPUs), equipped on many modern digital instruments \citep{Baan2004,sks,gmrt}.

Spectral Kurtosis (SK) is a higher-order statistical tool, applicable to accumulated Power Spectral Density (PSD) estimates, that has become increasingly popular over the last several decades in digital signal processing applications for automatic detection of non-Gaussian signals embedded in Gaussian background noise \citep{Dwyer,Ottonello,Vrabie,Antoni2006a,Antoni2006b,Antoni2007,Dion}. More recently, \citet{rfi} demonstrated that SK may be effective for automatic detection of RFI in astronomical signals, and proposed a simple design for a wide-band SK spectrometer based on its implementation and testing in a software correlator for a prototype instrument \citep{fst}.

As extensively developed in our preceding studies \citep{rfi,sk,sks,gsk}, what makes an SK spectrometer distinct from a traditional one is the fact that, for each frequency channel, it accumulates not only a set of $M$ instantaneous power spectral density (PSD) estimates, $S_1=\sum_{i=1}^MP_i$, but also the squared instantaneous spectral power, $S_2=\sum_{i=1}^MP_i^2$. A remarkable property of the spectral kurtosis estimator ($\esk$) computed from these parallel accumulations is that, in the case of a pure Gaussian time domain signal, its statistical expectation is unity for each frequency channel and time, no matter how the power level may vary in spectral shape or time. This property gives the \sk estimator the ability to discriminate signals that deviate from Gaussian time domain statistics against arbitrarily shaped astronomical backgrounds, thus making it sensitive to many man-made signals that produce unwanted RFI contamination of the astronomical signals of interest.

Moreover, the \sk estimator is expected to deviate below unity if the RFI signal is continuous or acting for more than half of the accumulation time, while it is expected to deviate above unity if the RFI signal has a less than $50\%$ duty-cycle.  Thus, the \sk estimator also provides the means to characterize the duration of the RFI signals relative to the instrumental accumulation time. Nevertheless, as shown by \citet{rfi}, the \sk estimator also deviates above unity in the case of a Gaussian transient lasting shorter than the integration time, which may pose a real-time decision challenge if such Gaussian transients represent wanted astronomical signals. However, as recently suggested by \citet{skt} in a comprehensive theoretical study, and experimentally demonstrated by \citet{jgr}, this sensitivity of the \sk estimator to transient astronomical signals, if properly taken advantage of, may provide an experimental means for obtaining accurate signal-to-noise ratio and duration estimates for both Gaussian and non-Gaussian transients, and thus extends the potential uses of the \sk spectrometer well-beyond its originally intended scope.

Developed for the Korean Solar Radio Burst Locator telescope \citep[KSRBL,][]{ksrbl}, the first hardware implementation of a wide-band spectrometer with SK capabilities, utilizing FPGA architecture, served as a demonstration \citep{sks} of the effectiveness of real-time RFI detection and excision via the SK algorithm.  In addition, the quantitative examination of the resulting high-precision probability density function (PDF) also revealed the need for a more accurate calculation of the theoretical RFI detection thresholds than the traditional, symmetrical standard deviation thresholds employed by \citet{rfi} in the original software-based SK spectrometer prototype.

Consequently, \citet{sk} proposed a more theoretically well-founded spectral kurtosis estimator (\sk), which applies to the case of a normally distributed time domain signal whose PSD estimates are obtained by means of a Fast Fourier Transform (FFT).  It is an unbiased estimator of the spectral variability, which is defined as the ratio between the variance and the squared mean of the PSD estimates.  For this particular case, \citet{sk} derived exact analytical expressions for the infinite series of statistical moments of the \sk estimator, and assigned to it a Pearson Type IV probability curve \citep{Pearson}, which was shown to be in very good agreement with numerically simulated \sk PDFs, as well as with the \sk distributions derived from direct experimental observations made with KSRBL \citep{sks}.

Given its conceptual simplicity, which makes it ideally suited for a real-time FPGA or GPU implementation, the original \sk spectrometer design proposed by \citet{rfi} has begun to be adopted as a standard component in the analysis pipelines of a new generation of radio telescopes designed not only by us, but also by other independent instrument design teams \citep{difx,allen,leap}.

However, one drawback of the SK spectrometer is the need to record both power and power-squared.  Despite rapidly increasing computational and storage capabilities and decreasing associated costs, the overhead brought by the doubling of real-time data volume and rates relative to that required by a traditional spectrometer design may still negatively impact the larger adoption of such instruments, especially in the case of radio interferometry arrays involving a large number of individual antennas. One solution to this is to compute \sk in the firmware and either record only the real-time SK flags, which can be represented by single bits, or even apply them in the firmware and dispense with recording power-squared entirely.  An alternative solution, which also broadens the use of the \sk estimator to existing spectrometers not specifically designed for SK, was developed by \citet{gsk}---the concept of a generalized \sk estimator that, with a trade-off of decreased temporal resolution, may be computed solely from the output of a standard spectrometer (one that provides only accumulated PSD estimates, $s_1=\Sigma_{j=1}^{N}P_j$).

As defined by \citet{gsk} in terms of the accumulations $S_1=\Sigma_{i=1}^M{(s_1)}_i$ and $S_2=\Sigma_{i=1}^M{(s_1)}^2_i$, and the shape parameter $d$, the generalized spectral estimator is
\begin{eqnarray}
\label{GSK}
\widehat{SK}=\frac{M N d+1}{M-1}\Big{(}\frac{MS_2}{S_1^2}-1\Big{)},
\end{eqnarray}
where $N$ is the number of internal accumulations performed to produce the instrumental output $s_1$, and $M$ is the number of such consecutive accumulations used to compute the above sums $S_1$ and $S_2$.  This generalized form of \sk has unity expectation not only in the particular case of accumulations involving raw PSD estimates obtained via FFT, which have a Gamma probability distribution of shape parameter $d=1$ (exponential distribution), but also for PSD estimates characterized by Gamma probability distributions of arbitrary shape parameters $d$, particularly the PSD estimates obtained by means of narrow-band filtering of wide-band time domain signals, which obey a chi-square statistics corresponding to a Gamma distribution with shape parameter $d=0.5$. Therefore, unlike its original counterpart, the generalized \sk estimator defined by Equation \ref{GSK} is no longer limited to a particular instrument design or method used to obtain the PSD estimates.

It may be helpful to give some specific use cases: (i) For a spectrometer specifically designed for SK, which produces $S_1$ and $S_2$ in firmware, one would use $N=1$, $d=1$ in Equation~\ref{GSK}. (ii) For a typical FFT-based spectrometer that outputs only accumulated power $s_1$, one would set $N$ to the number of on-board accumulations, and then perform offline accumulation of $S_1$ and $S_2$, and use $d=1$. (iii) For a spectrometer that outputs only a band-limited time-domain signal, one would set $N$ to the number of on-board accumulations of data samples, perform offline accumulation of $S_1$ and $S_2$, and use $d=0.5$. (iv) Finally, one may sometimes have data from a spectrometer whose internal characteristics, and hence number of on-board accumulations, are not known.  In this case, it is often possible to calculate the partial estimate for large $M$, $\widehat{SK}'=(MS_2/S_1^2-1)$ and observe its value in RFI-free regions, to estimate $Nd \approx (MS_2/S_1^2-1)^{-1}$.

Although Equation~\ref{GSK} suffices to calculate the expected \sk for different use cases, it is also necessary to know the \sk PDF for each case, so that thresholds can be computed beyond which one can be confident, at some probability level, that the signal is non-Gaussian. Following the same mathematical framework used to obtain the \sk PDF approximation corresponding to the particular case $\{N=1, d=1\}$ \citep{sk}, \citet{gsk} derived the infinite series of statistical moments of the generalized \sk estimator defined by Equation~\ref{GSK} and demonstrated that, for different regions of the parameter space $\{M, Nd\}$, the \sk PDF may be approximated by one of a Pearson Type I, Type III, or Type IV probability distribution, any of which can provide known probability of false alarm (PFA) detection thresholds accurate enough for practical applications. In Appendix \ref{code} we provide an example procedure written in Interactive Data Language (IDL) that may be used to compute such thresholds.

Just as KSRBL \citep{ksrbl} was the first implementation of an SK spectrometer, we report here on the first implementation of SK in a multi-antenna, interferometric array, the Expanded Owens Valley Solar Array (EOVSA).  In doing so, we extend the previous work by demonstrating the use of both the traditional \sk estimator and the generalized \sk estimator for data recorded with the system.  In \S\ref{design} we describe the design of the EOVSA correlator, which implements the SK function.  In \S\ref{tests} we discuss specific tests that demonstrate the performance.  We conclude with some general remarks in the final \S\ref{conclusion}.

\section{EOVSA SK Correlator Design\label{design}}

Following our success with the KSRBL spectrometer design described by \citet{ksrbl}, we have implemented a similar design in the Fourier transform engine (F-engine) of the 16-antenna, dual-polarization FX correlator for the Expanded Owens Valley Solar Array (EOVSA).  EOVSA is a solar-dedicated array of 13 small 2.1-m-diameter dishes equipped with broadband (1--18 GHz) front-end systems with crossed-linear-polarization feeds.  A pair of 20~GHz optical links transmit the entire RF band from the two polarizations to a central control building, where they are downconverted via a tunable LO to sample an instantaneous bandwidth of 600~MHz. The system is capable of tuning to a new IF band in under 1~ms, and the correlator is designed to accumulate for 19~ms, thus providing a 20~ms sample time.  The frequency-agile system is capable of tuning in arbitrary order, and typically tunes to 50 IF frequencies in a fixed 1~s cycle.  The correlator also accepts input from two 27-m antennas equipped with He-cooled receivers, which are used for observation of celestial sources for calibration of the small dishes.

\begin{figure}[h]
\label{design_fig}
\begin{center}
\includegraphics[scale=0.80]{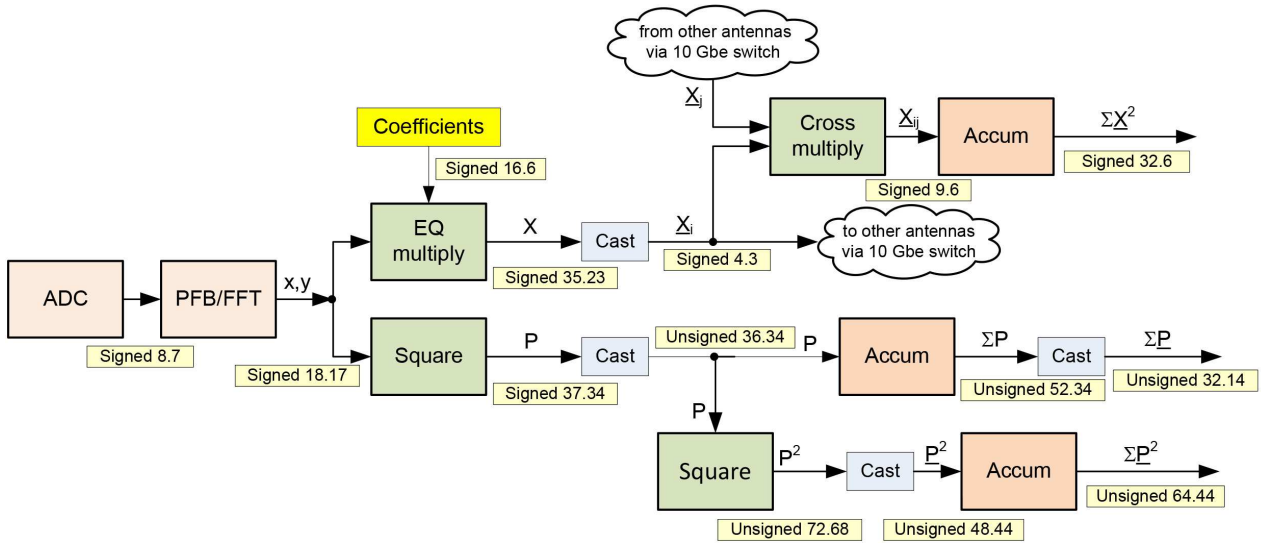}\caption{\it Block diagram of EOVSA correlator.  Quantities with no bit-truncation are shown without an underbar, while quantities with potential truncation are shown with an underbar.  Truncation is performed via the "Cast" blocks, which change the output bit-width for practical purposes of data volume and compatibility with standard computational data types, e.g. 32-bit data for accumulated power ($\sum \underline{P}$) and cross-power ($\sum \underline{X}^2$), and 64-bit data for power-squared ($\sum \underline{P}^2$).}
\end{center}
\end{figure}

The digital part of the system, comprising the ADCs, the F-engine, and the cross-correlation engine (X-engine), is based on the hardware and firmware of the Collaboration for Astronomical Signal Processing and Electronics Research (CASPER), and is implemented on second-generation Reconfigurable Open Architecture Computing Hardware (ROACH-2) boards equipped with first-generation, dual-input Karoo Array Telescope analog-to-digital converter (KatADC) boards.  Each KatADC board accepts input from the two polarizations of a single antenna, and each ROACH-2 has two such boards, for four inputs in total.  The entire 16-antenna correlator thus requires 8 ROACH-2 boards.  The firmware for each ROACH-2 is running four parallel, identical F-engines and a single X-engine that handles 1/8th of the total IF bandwidth on a single Xilinx Virtex-6 Field Programmable Gate Array (FPGA).  The remaining 7/8ths of the band is sent via 10-gigabit ethernet packets to the other 7 ROACH-2 boards.  The F-engine and X-engine data are accumulated over 19 ms on each board, and then sent via 10-gigabit ethernet to a processing computer. Figure~\ref{design_fig} shows a simplified block diagram of the digital hardware.  At present (August 2016), the functional design of the digital correlator is final, but the design is operating at an FPGA clock speed of 200~MHz, lower than the design goal of 300~MHz.  Small design tweaks are being applied in order to bring the timing to the design goal of 300~MHz, which is expected soon. Meanwhile, the tests described in the next section were taken with the lower clock speed, which limits the effective IF bandwidth to 400~MHz and causes an overlap of the lower half of the band, so that it is effectively double-sideband.  The upper half of the 400~MHz band is the preferred single-sideband.  However, these complications in no way affect the operation of the system for the purposes of evaluating SK performance.

Figure~\ref{design_fig} gives the bit-width of the digital representation of numerical values at various points in the design. To properly determine \sk from the quantities output by the EOVSA correlator, care must be taken to keep the precision of the approximate output values ($\sum \underline{P}$) and ($\sum \underline{P}^2$) as high as possible without overflow.  The critical locations where inappropriate truncation could occur are the ``Cast" block just after the accumulator for the accumulated power, and the ``Cast" block just before the accumulator for the individual power-squared samples.  Since these blocks merely take the most significant bits in both cases, this amounts to ensuring that the power be as high as possible without overflowing.  There are two adjustments that must act together to keep the power at the desired level.  First, the overall analog system gain must be adjusted to keep the digitized ADC signal in an appropriate range, which anyway is needed for optimum performance of the digital system, and second, a quantity called the FFT shift can be manipulated to keep the output of the FFT in the appropriate range for good SK performance.  EOVSA's 4096 frequency channels require a 13-stage FFT, and at each stage it is possible with the CASPER firmware to shift the output by one bit to lower precision.  With no shifts at all, FFT overflows are highly likely, so shifting the output of some FFT stages is often needed.  Each shift of a stage results in the $x,y$ quantities being reduced by a factor of 2, and after squaring to form the power $P$, the power reduces by a factor of 4 (and power-squared by a factor of 16).  The procedure we use to determine the optimum FFT shift is to simply try different values and form the \sk estimator, which will be unity for non-RFI affected channels for a fairly wide range of FFT shift values, but will differ from unity when the FFT shift is too small, so that $P$ starts to overflow, or too large, so that $P^2$ starts to fall to low precision.

In practice, EOVSA's design provides such a high precision (64 bits) for $\sum P^2$ that \sk is well behaved over a wide range of FFT shifts.  When only the first 3 or fewer stages are shifted, the \sk values fall below unity for some non-RFI channels, but they are well behaved when at least 4, and up to 10, stages are shifted.  This range of 7 bits in $x,y$ corresponds to a range of 14 bits in $\sum P$ and 28 bits in $\sum P^2$, yet the SK performance is unchanged over this range. For our operational setting, we have settled on shifting the first 5 stages, which gives some headroom in case of flares while keeping the precision high.  Although not relevant to SK operation, we note that once the FFT shift is set, it is a more delicate matter to keep the $X$ data in range due to its cast to only 4 bits prior to cross-correlation, via the ``Cast" block after the equalizer block (labeled ``EQ multiply" in Figure~\ref{design_fig}).  This is done by applying a $34\times128$ array of equalizer coefficients (yellow block in the figure), which can vary by band and channel, with 128 coefficients per band so that each applies to 32 of the 4096 channels in each band.  The equalizer coefficients are adjusted by a calibration procedure that seeks to make the rms of the 4-bit autocorrelations lie in the range 2 to 3.

\section{EOVSA SK Performance Tests\label{tests}}

As described in \S\ref{design}, the normal mode of operation of EOVSA consists of cycling a user-defined tuning sequence composed from a set of 34 predefined 400~MHz-wide IF bands (will be 600~MHz-wide when timing is met), some of them repeated, such that the entire (1--18~GHz) frequency range is covered in 1~s by 50 independent tunings, resulting in 19-ms-long power and power-square accumulations ($M=1792$ raw FFT spectra), which are divided into 4096 frequency channels ($\sim0.098$~MHz resolution). If the real-time RFI-excision algorithm is activated, the high-frequency-resolution $S_1$ and $S_2$ accumulations are then used to compute \sk estimators according to equation \ref{GSK} ($N=1, d=1$). \sk values that fall outside the thresholds are interpreted as RFI-contaminated channels, which are excluded from a subsequent frequency integration performed in order to produce the final 1~s time-resolution dynamic spectrum characterized by, typically,  $\sim500$ ``science channels" (with variable frequency resolution up to $\sim40$~MHz). However, for the purpose of evaluating the performance of the hardware-embedded SK engine, we first analyze a data segment obtained by dwelling on a fixed 400~MHz IF band (9.6--10~GHz), which happens to be virtually free of RFI.

The results of statistical analysis displayed in Table~\ref{sk_table}, which was performed for the 13 EOVSA antennas, 2 linear polarizations labeled XX and YY, 4096 frequency channels, and 1000 contiguous accumulations of $M=1792$ raw FFT spectra each, demonstrates a full agreement of results with the theoretical expectations of \citet{sk,gsk}. Indeed, despite the variations of bandpass shapes of individual antennas and polarizations, the distribution of the observed 4,096,000/antenna/polarization \sk samples displayed in Table~\ref{sk_table} have frequency- and polarization-independent means close to unity, and sample variances consistent with the theoretical expected variance $\sigma_{\widehat{SK}}=2/\sqrt{M}=0.047245$ corresponding to $M=1792$. The accuracy of the Pearson Type IV approximation of the \sk PDF derived by \citet{sk} is consistently confirmed by the observed percentages of flagged data by a pair of asymmetric thresholds, $\esk\le0.87145$ and $\esk\ge1.15800$, chosen to correspond to an intended $0.13499\%$ PFA.

\begin{wstable}[h]
\caption{{\it EOVSA SK statistical analysis of RFI-free data.} The sample \sk means and variances are to be compared with the theoretical expectations given in the first line of the table. The percentages of flagged frequency-time bins below and above unity are to be compared with the expected equal probabilities of false alarm corresponding to the asymmetric RFI detection thresholds, i.e. PFA$(\esk\le0.871)=0.13499\%$ and PFA$(\esk\ge1.158)=0.13499\%$, respectively.}
\begin{tabular}{@{}cccccc@{}} \toprule
& $\langle\widehat{SK}\rangle$& $\sigma_{\esk}$ & $\le0.87145$ & $\ge1.15800$\\
   & & & \%&\% \\ \colrule
Expected $\rightarrow$  &1.00000 &0.04725 &0.13499 &0.13499\\ \colrule
Antenna  1 XX &0.99998 &0.04724 &0.14068 &0.13926\\
Antenna  1 YY &1.00014 &0.04793 &0.13168 &0.16574\\
Antenna  2 XX &1.00007 &0.04751 &0.13638 &0.16306\\
Antenna  2 YY &0.99992 &0.04727 &0.15430 &0.13854\\
Antenna  3 XX &1.00025 &0.04963 &0.13343 &0.16271\\
Antenna  3 YY &1.00017 &0.04732 &0.13527 &0.15209\\
Antenna  4 XX &1.00011 &0.04813 &0.13486 &0.16325\\
Antenna  4 YY &0.99992 &0.04771 &0.16015 &0.13854\\
Antenna  5 XX &1.00027 &0.04972 &0.13360 &0.16281\\
Antenna  5 YY &0.99993 &0.04739 &0.16119 &0.13515\\
Antenna  6 XX &1.00022 &0.04961 &0.13567 &0.16111\\
Antenna  6 YY &0.99997 &0.04755 &0.16040 &0.13336\\
Antenna  7 XX &1.00000 &0.04725 &0.13458 &0.14543\\
Antenna  7 YY &1.00001 &0.04725 &0.14594 &0.13790\\
Antenna  8 XX &1.00019 &0.04942 &0.13631 &0.16399\\
Antenna  8 YY &0.99994 &0.04763 &0.15964 &0.14122\\
Antenna  9 XX &1.00026 &0.04963 &0.13422 &0.16298\\
Antenna  9 YY &0.99994 &0.04748 &0.15374 &0.13975\\
Antenna 10 XX &1.00015 &0.04875 &0.13402 &0.15949\\
Antenna 10 YY &1.00004 &0.04739 &0.13446 &0.15563\\
Antenna 11 XX &1.00000 &0.04741 &0.15745 &0.13581\\
Antenna 11 YY &0.99999 &0.04730 &0.14968 &0.13987\\
Antenna 12 XX &0.99994 &0.04725 &0.14331 &0.14117\\
Antenna 12 YY &1.00003 &0.04726 &0.14523 &0.13640\\
Antenna 13 XX &1.00002 &0.04721 &0.13070 &0.14012\\
Antenna 13 YY &0.99994 &0.04732 &0.15384 &0.13815\\
\botrule
\end{tabular}

\label{sk_table}
\end{wstable}

In the next performance test, we tuned the EOVSA instrument to a fixed 400~MHz IF band prone to RFI contamination (3.6--4.0~GHz), and then commanded the antennas to point from an on-Sun position to off-Sun, to simulate a change in total power level that might be seen during the evolution of a solar burst. Although $S_1$ thus varies by about a factor of 3, the value of the \sk estimator is expected to be unchanged due to its strict lack of dependence on changes of the background power level.

\begin{figure}[h]
\begin{center}
\includegraphics[scale=0.78]{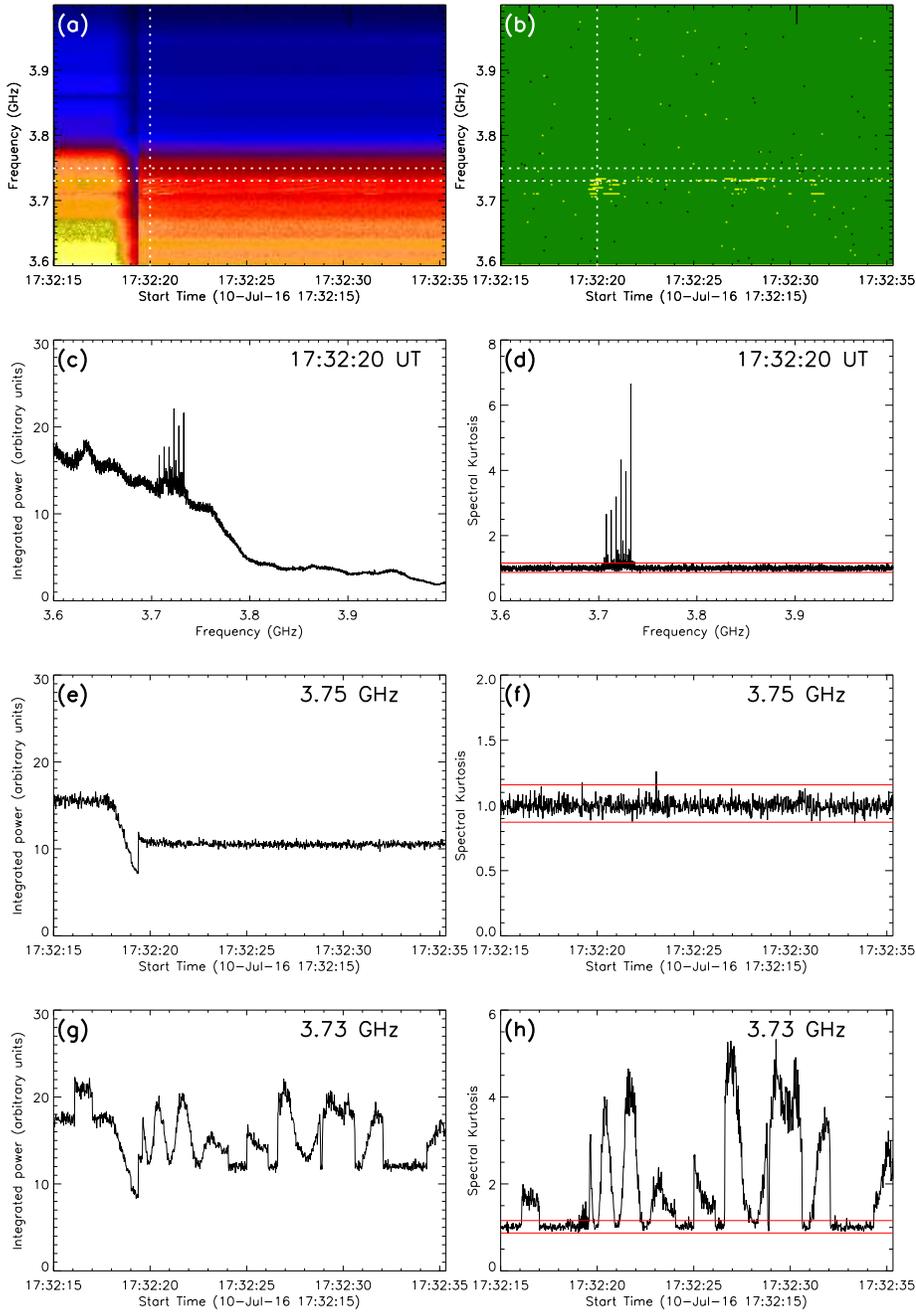}
\end{center}
\caption{{\it Fixed IF band SK analysis (3.6--4 GHz).} a) Antenna 1 dynamic power spectrum for XX polarization, with $\Delta f=0.097$~MHz; $\Delta t=0.02$~s; $M=1792$, during an on-to-off-center Sun track. The vertical and horizontal dotted lines indicate one time and two frequency channels selected for the line-plots in subsequent panels. b) Color-coded dynamic spectrum of the corresponding \sk flags. A green, uniform background indicates \sk within the computed thresholds. Black and yellow colors indicate the presumably RFI-contaminated time-frequency bins characterized by $\esk\le0.87145$ and $\esk\ge1.15800$, respectively, which represent the theoretical $M=1792$ RFI thresholds corresponding to an intended $0.13499\%$ PFA. c) Bandpass accumulated power spectrum corresponding to the vertical dotted-line marker shown in a and b. d) \sk spectrum corresponding to the same selected time slot as in panel c, overplotted with the thresholds in red. e) Time evolution of the integrated power at 3.75~GHz. f) Time evolution of \sk corresponding to panel e, overplotted with the thresholds in red. g) same as e, for the RFI-contaminated frequency 3.73~GHz. h) Time evolution of \sk at 3.73~GHz. The varying $\esk\ge1.15800$ values indicate intermittent RFI contamination with evolving $<50\%$ duty-cycles relative to the integration time.
}
\label{slew}
\end{figure}

Figure~\ref{slew}a displays the XX-polarization dynamic power spectrum for one of the antennas obtained in the $3.6-4$~GHz frequency range, with frequency resolution $\Delta f=0.097$~MHz, and time resolution $\Delta t=0.02$~s, the latter corresponding to an accumulation length of $M=1792$ raw FFT spectra.  The change of the background on a time scale of about 2~s during the on-off Sun transition is evident.  The sharp transition to higher power level at about 17:32:19~UT is due to the automatic gain control switching out 2~dB of attenuation to compensate for the decreased radio frequency (RF) level, which affords us the opportunity to characterize the SK performance for both gradual and abrupt power-level changes.  Also apparent from Figure~\ref{slew}a is the nonuniform bandpass shape reflecting the (temporary) overlap of the lower half of the band mentioned in \S\ref{design}.  Relatively faint RFI contamination may be observed superposed on the background spectrum in some time--frequency bins that are crossed by the vertical dotted line and one of the two horizontal dotted lines that mark one instant of time and two frequency channels selected for a more detailed analysis presented in the subsequent panels of Figure~\ref{slew}.

Figure~\ref{slew}b displays a color-coded dynamic spectrum of \sk flags corresponding to the power spectrum shown in panel \ref{slew}a. The green uniform background indicates the RFI-free time-frequency bins where \sk is within the calculated thresholds. The black and yellow regions indicate the presumably RFI-contaminated time-frequency bins characterized by $\esk\le0.87145$ and $\esk\ge1.15800$, respectively, which represent the theoretical $M=1792$ RFI thresholds corresponding to an intended $0.13499\%$ PFA. The clustered yellow $\esk\ge1.15800$ flags 
clearly seen in the \sk flag spectrum indicate the presence of RFI signals characterized by $<50\%$ duty-cycle relative to the integration time, while the more uniformly scattered $\esk\le0.87145$ and $\esk\ge1.15800$ flags mostly correspond to RFI-free \sk statistical fluctuations, as we confirm by comparing the actual percentage of the $\esk\le0.87145$ flags, i.e. $0.20093\%$, with the theoretically expected $0.13499\%$ PFA, as opposed to the statistically significantly larger $0.50789\%$ percentage of $\esk\ge1.15800$ flags.

Figure~\ref{slew}c displays the accumulated power spectrum corresponding to the vertical dotted-line marker shown in Figure~\ref{slew}a and \ref{slew}b. Superposed on the variable-shape bandpass spectrum, relatively strong RFI contamination is clearly seen in the $3.71-3.74$~GHz frequency range. The corresponding \sk spectrum displayed in Figure~\ref{slew}d generally reveals a flat unity mean \sk distribution bounded by the computed $0.13499\%$ PFA thresholds (horizontal red lines). The RFI contamination in the $3.71-3.74$~GHz range is clearly flagged by larger than unity \sk values, which are indicative of RFI contamination with $<50\%$ duty-cycle relative to the integration time \citep{skt,jgr}.

Figure~\ref{slew}e displays the evolution of the integrated power at the upper of the two horizontal dotted-line markers, corresponding to 3.75~GHz. The progressive drop of intensity from on- to off-Sun levels is evident. The abrupt increase of the power level at 17:32:19 UT due to the automatic gain control is also evident. The corresponding \sk lightcurve displayed in Figure~\ref{slew}f, which fluctuates within the statistical $0.13499\%$ PFA thresholds (horizontal red lines), is remarkably independent of the variations seen in the background power level, which indicates no RFI contamination in the selected frequency channel.

Figure~\ref{slew}g displays the evolution of the integrated power at the lower horizontal marker, 3.73~GHz, which, in contrast with the time evolution seen in the almost adjacent 3.75~GHz channel, features strong signal variations superimposed on a background level similar to the adjacent RFI-free frequency channels. The corresponding \sk evolution at 3.73~GHz is shown in Figure~\ref{slew}h. The varying $\esk\ge1.15800$  values indicate intermittent RFI contamination with evolving $<50\%$ duty-cycles relative to the integration time. Nevertheless, the \sk time profile clearly demonstrates the ability of the SK-based RFI excision algorithm to adaptively select contiguous RFI-free time segments suitable for science investigations, as opposed to a non-adaptive procedure that would discard entirely such heavily RFI-contaminated channels.

\begin{figure}[h]
\begin{center}
\includegraphics[scale=1.0]{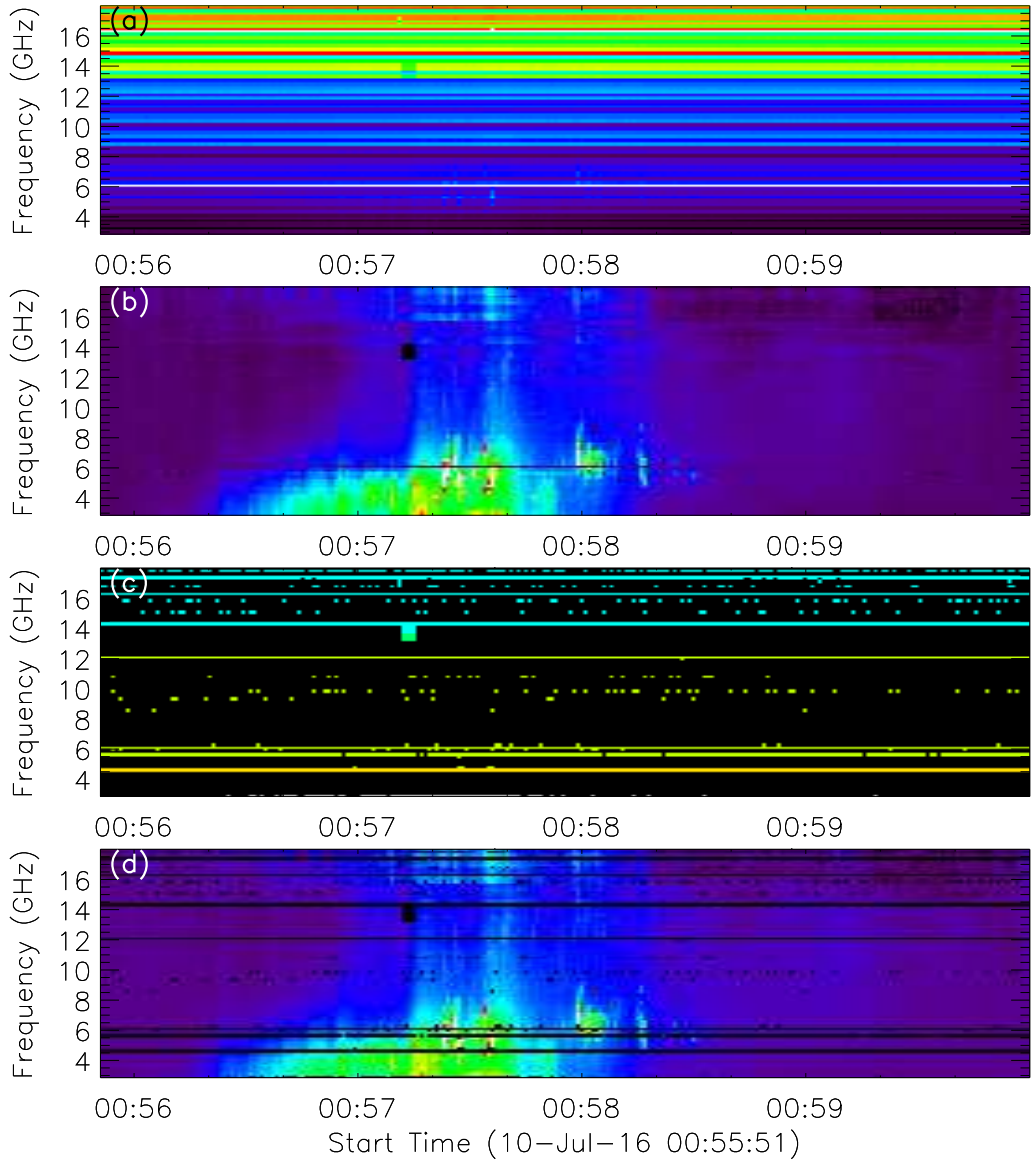}
\end{center}
\caption{{\it \sk analysis based on EOVSA data recorded during the GOES C8.6 flare of 2016-07-10 00:53:00~UT in the $[2.86-17.98]$~GHz frequency range.} a) Raw (uncalibrated, not background-subtracted) $S_1$ spectrum. b) Calibrated and background-subtracted power spectrum that reveals a more or less smooth flare envelope, on top of which a set of clustered fine spectral features are clearly visible. The RFI contaminated channel around 6 GHz is also visible in this panel. c) RFI flag spectrum corresponding to variable $1+6/\sqrt{M}$ thresholds, as explained in the main text. A much larger percent of RFI contaminated channel than suggested by the plots shown in a and b is revealed. d) RFI-flagged dynamic spectrum. Remarkably, almost none of the observed fine structures are flagged, which confirms their solar origin.}
\label{flare}
\end{figure}

In a final test of the SK performance, in Figure~\ref{flare} we present a data segment recorded by the EOVSA instrument during the GOES C8.6 flare of 2016-07-10 00:53:00~UT. However, since at the time of this event the real-time RFI excision algorithm was turned off, we do not have the full-resolution SK data, but rather only $S_1$, $S_2$ and $M$ accumulated over the wider, and variable-width, science channels. The dynamic spectrum ranges from 2.86--17.98~GHz, and the time resolution is 1~s, which, as explained in \S2, are a result of the tuning cycle of 1~s and a frequency-domain integration resulting in 134 science channels. Moreover, prior to the science-band integration, the $S_1$ and $S_2$ accumulations corresponding to any repeated IF bands during the 1~s cycle were added together. As the result of these operations, the reduced-resolution $S_1$ and $S_2$ corresponding to the science channels are characterized by different accumulation lengths $M$. For the data set analyzed in Figure~\ref{flare}, the number of accumulations range from $M=164,864=90\times M_0$ for the narrower science bands at lower frequencies, to $M=1,924,608=1074\times M_0$ for the broader (and repeated) science bands at higher frequencies, where $M_0=1792$ is the output accumulation length of the EOVSA correlator.

Figure~\ref{flare}a displays the raw $S_1$ spectrum that results from this procedure. Except for a strong RFI-contaminated channel, the raw solar spectrum appears almost featureless due uncalibrated, frequency-dependent gain variations, and to the strong gradient of the quiet Sun signal, which reaches more than 500 solar flux units (sfu; 1 sfu = $10^{-22}$ W m$^{-2}$ Hz$^{-1}$ = 10000 Jy) at high frequencies. Only after subtracting this background signal and applying calibration does the $<50$~sfu solar burst become clearly visible. The calibrated power spectrum derived from $S_1$, which is displayed in Figure \ref{flare}b, reveals the flare spectrum composed of a more or less smooth envelope, on top of which a set of clustered fine spectral features are clearly visible. However, except for the clearly RFI-contaminated channel near 6~GHz, it is not immediately evident whether other frequency-time bins are affected by RFI. Although the frequency-time integration is an irreversible process, the availability of the $S_1$, $S_2$ and $M$ still allows the computation of an \sk estimator for each time-frequency bin, and thus discrimination of those bins affected by RFI signals.

Similar to the procedure used to obtain the RFI flags displayed in Figure~\ref{slew}b,  we show in Figure~\ref{flare}c the RFI flag dynamic spectrum obtained  by assigning a flat unity value to all spectral bins characterized by \sk that fluctuate within the $1\pm6/\sqrt{M}$ range (black background), while saturating to these $M$-dependent values all spectral bins characterized by \sk values ranging below or above these limits.  The choice of these $1\pm3\sigma_{\esk}$ symmetric thresholds, instead of the exact $0.13499\%$ PFA thresholds used in Figure \ref{slew}b, is motivated by the extremely large $M$ values, for which the Gaussian approximation of the true \sk PDF is fully justified \citep{sk}. Figure~\ref{flare}c reveals that the RFI contamination is stronger than suggested by the visual analysis of the dynamic spectra shown in Figure~\ref{flare}a,b. Indeed, a quantitative analysis demonstrates that $12.57\%$ of the time-frequency bins feature \sk values above the upper $1+6/\sqrt{M}$ threshold, while no spectral bin is found to have $\esk\le 1-6/\sqrt{M}$.

Remarkably, the \sk flagged power spectrum shown in Figure \ref{flare}d, demonstrates that, with a few noticeable exceptions discussed below, the vast majority of the \sk flags are not related to the flare evolution. This demonstrates the ability of the \sk estimator to automatically discriminate RFI contamination against natural spectral features, even in the case of fine spectral structures as those seen superposed on the flare background. Nevertheless, a few flagged spectral bins coincide with some of the fine spectral features, which may be just simple coincidence, or the result of a fast evolution of the such spectral features at a time scale shorter than the integration time, as theoretically expected by \citet{skt} and experimentally demonstrated by \citet{jgr} using data recorded by an EOVSA prototype instrument during a flare featuring solar radio spikes. Therefore, aside from these few exceptions, the \sk flags may be safely attributed to RFI contamination, and corresponding data may be safely discarded from scientific analysis of the event, particularly imaging when available, given the fact that EOVSA is capable designed to produce a full Sun image for very time-frequency bin of the dynamic spectrum.

We also find noteworthy the fact that none of the \sk flags are found to correspond to \sk values less than unity, which would be the case for an RFI signal lasting longer than the instrumental integration time \citep{skt,jgr}, which in the particular case of the EOVSA tuning sequence may range anywhere between 20~ms (for non-repeated bands) and 60~ms (for bands repeated 3 times). However, rather than asserting that no such RFI signals were mixed in the data, we point out that, even for a purely continuous, narrowband RFI signal, which would be detected as less than unity by the real-time RFI excision algorithm, the integration across frequency channels to form the wider-band science-frequency channels effectively reduces the contribution of RFI to the integrated signal to a percentage much smaller $50\%$, and thus results in \sk values larger than unity.

We conclude this analysis by pointing out that, had the automatic RFI excision algorithm been active at the time of the flare, most of the $12.57\%$ bins contaminated by RFI could have been saved from being fully discarded, and the scientific return of the instrument significantly increased.

\section{Discussion and Conclusion\label{conclusion}}

\begin{figure}[h]
\begin{center}
\includegraphics[scale=0.9]{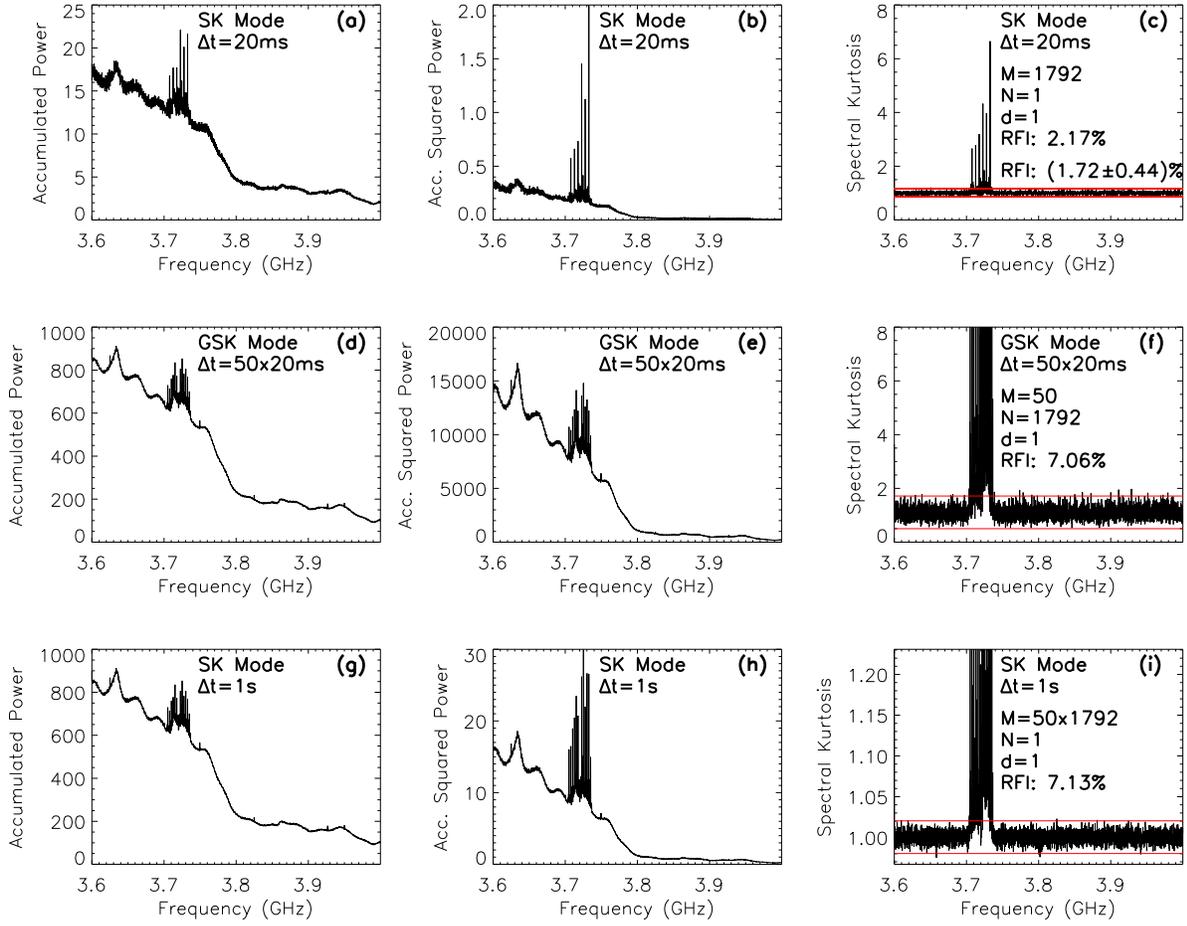}
\end{center}
\caption{{\it \sk analysis comparing use of EOVSA data in SK mode versus use as a traditional spectrometer in generalized SK mode.} a) The 20~ms accumulated power spectrum from Figure~\ref{slew}c. b) The accompanying 20~ms accumulated power-squared spectrum output. c) Computed \sk spectrum (black), and 0.13499\% PFA threshold levels (red horizontal lines) from Figure~\ref{slew}d. d) Sum of 50 consecutive 20-ms power accumulations covering 1~s of data, starting with the time frame illustrated on the top panels. e) Sum of the squares of the same 50 consecutive 20~ms power accumulations. f) Computed generalized \sk spectrum for the band (black) and the threshold levels (red horizontal lines). The same vertical scale as in panel (c) has been imposed to facilitate  comparison of the \sk and generalized \sk fluctuations. g) The same 1~s accumulated power spectrum shown in panel (d). h) The 1~s accumulated power-squared spectrum directly obtained by summing the 50 consecutive 20~ms power squared accumulations natively produced by the EOVSA spectrometer. i) The \sk spectrum computed based on the 1~s power and power-squared accumulations. The vertical scale of the \sk plot shown in panel (i) has been zoomed-in by a $\sim30:1$ factor to illustrate the practical statistical equivalence of the the 1~s \sk and generalized \sk spectra, despite the $\sim30:1$ ratio of their absolute statistical fluctuations. In each panel on the right column we indicate the actual $M$, $N$ and $d$ values entered in Equation \ref{GSK} to compute the \sk spectra. The percentages of RFI flagged channels are displayed for each \sk plot in the right column. In panel (c), the mean and standard deviation of the RFI percentages corresponding to each of the 50 consecutive time frames are also displayed.}
\label{gsk_fig}
\end{figure}

Although not a test of EOVSA's SK performance per se, we can use EOVSA data to better illustrate in Figure ~\ref{gsk_fig} the contrast between using its output as an SK spectrometer versus its output as a traditional spectrometer lacking specially designed SK capabilities.

We already showed an example in Figure~\ref{slew}d of the \sk estimator for a single, 20~ms snapshot, where the lower and upper thresholds for $M=1792$, $N=1$ and $d=1$ are 0.87145 and 1.15800, respectively.  Figure~\ref{gsk_fig}a,c are the same as Figure~\ref{slew}c,d, to aid comparison. In addition, Figure~\ref{gsk_fig}b also shows the power-squared accumulation corresponding to the same 20~ms time slot. As indicated in the Figure~\ref{gsk_fig}c inset, the 0.13499\% PFA \sk thresholds (red horizontal lines) flag $2.17\%$ of the 4096 frequency channels as being contaminated by RFI. For later comparison, the inset also gives the mean and standard deviation of RFI occupancy ($1.72\pm0.44$\%) for 50 consecutive 20~ms samples. When compared with the much lower 0.13499\% PFA corresponding to the RFI thresholds, the $0.44\%$ standard deviation corresponding to the analyzed sequence clearly indicates a true change in RFI occupancy from one accumulation to another. We consider this finding to be fully consistent with the observed, larger-than-unity \sk values, which can only arise as the result of RFI signals that change their dynamics at time scales shorter than the integration time \citep{rfi, skt}, and thus may not be necessarily active, or detectable, in all consecutive accumulations.

Now imagine that the EOVSA correlator does not provide power-squared information, but instead provides only accumulated power, $s_1$.  We can still calculate \sk using the generalized \sk formula (Equation \ref{GSK}), where $S_1$ and $S_2$ are accumulated from sums of $s_1$ and $s_1^2$ over some time interval.  The middle row of Figure~\ref{gsk_fig} illustrates this approach for a sum of 50 consecutive spectral samples (1~s) of the same data as in Figure~\ref{slew}a, where now the generalized \sk is computed using $M=50$, $N=1792$, and $d=1$.  Figure~\ref{gsk_fig}d displays the 1~s accumulated power, while Figure~\ref{gsk_fig}e displays the sum of the squares of the same sequence of accumulated power spectra. Based on these inputs, we calculate the generalized \sk spectrum displayed in Figure~\ref{gsk_fig}f, which, despite being computed from a much larger data set, is affected by much larger statistical fluctuations than those corresponding to the \sk plot displayed in Figure~\ref{gsk_fig}c. However, this is an expected result \citep{gsk}, since the statistical fluctuations affecting the generalized \sk estimator are ultimately dictated by the much smaller number of terms entering the outer sum, i.e $M=50$ versus $M=1792$.

Nevertheless, the confidence level of the RFI flags is not dictated by the absolute value of the \sk statistical fluctuations, but by the accuracy with which the RFI threshold are computed for a predefined PFA \citep{sk, gsk}. Indeed, the lower and upper thresholds, shown by red horizontal lines in Figure~\ref{gsk_fig}c, which are now 0.50077 and 1.71615, respectively, do not appear to be crossed by more than the expected 0.13499\% PFA of \sk values, except for the clearly RFI contaminated spectral region flagged in both (c) and (f) panels.

From this perspective, the significantly larger percentage of RFI occupancy, $7.06\%$, produced by the generalized \sk analysis over the same data segment on which the sequence of 50 \sk estimators indicates only $1.72\pm0.44\%$ RFI occupancy, cannot be attributed to a less accurate generalized \sk estimation, but to the dynamical nature of the RFI signals themselves, which during a longer time window contaminate more frequency channels than the number contaminated in a single 20~ms period. This conclusion is also supported by much larger generalized \sk values, ($\esk_{max}\sim86.5$ vs. $\esk_{max}\sim2.9$), which is a immediate consequence of the RFI signals not being continuously active in each frequency bin \citep{skt}.

To conclude our comparison between the \sk and generalized \sk approaches, we display on the bottom row of Figure~\ref{gsk_fig} the result of the \sk analysis involving both outputs of the EOVSA spectrometer over the entire 1~s time window, from which accumulations of $M=50\times1792$ power and power-squared samples may be generated (panels g and h, respectively). To facilitate a direct comparison of the \sk spectra displayed on panels (f) and (i), the vertical scale of the \sk plot shown in panel (i) has been zoomed-in by a $\sim30:1$ factor, which demonstrates their statistical similarity. Remarkably, when compared with the 0.13499\% PFA, the difference between the $7.13\%$ and  $7.06\%$ percentages of RFI flags is not statistically significant, which quantitatively demonstrates the statistical equivalence of the \sk and generalized \sk estimators in terms of their RFI sensitivity, when computed based on data spanning the same time window. Although, in the light of this comparison, it may be concluded that the same RFI detection performance is obtained by generalized \sk without the need of doubling the data output of a spectrometer, our comparison also demonstrates that this may come at the cost of flagging and discarding potentially good data (of order 5\% in this example), and on a time scale much larger than the native resolution of the instrument. Moreover, if the astronomical signal of interest is expected to vary on a time scale shorter than the analysis time window, as noted in \citet{rfi} \citep[see also][]{skt,jgr}, flagging of such astronomical transients will occur. Therefore, all these aspects should be considered when designing a new instrument.

In any case, as we already pointed out in the introductory section, a native \sk design may extend the potential uses of the \sk spectrometer well-beyond its originally intended scope by providing an experimental means for obtaining accurate signal-to-noise-ratio and duration estimates for both Gaussian and non-Gaussian transients \citep{{skt, jgr}}. Moreover, due to the additive property of both its outputs in both frequency and time, a native \sk spectrometer may also offer versatile means of employing, in an automated manner, the concept of multi-scale \sk analysis \citep{sks}, not only for improving the sensitivity of RFI excision \citep{sks}, but also for the purpose of automatic detection and classification of the statistical nature of signals \citep{skt}. A direct illustration of the multi-scale \sk analysis concept is provided by the right column of Figure~\ref{gsk_fig}, where the maximum values of \sk vary from $\esk_{max}\sim2.9$ in panel (c), to $\esk_{max}\sim86.5$ in panel (f), and $\esk_{max}\sim3.1$ in panel (i), as different duty cycles of RFI are generated for the same data.

We have described the hardware implementation of a correlator for EOVSA whose F-engine is designed to accumulate the sums of power and power-squared needed to compute the \sk estimator defined in Equation~\ref{GSK}.  Using actual data from EOVSA, we have demonstrated that it performs exactly as expected in a series of stringent tests, for RFI-free data (Table~\ref{sk_table}), data with RFI (Figure~\ref{slew}), and data containing both RFI and solar radio bursts (Figure~\ref{flare}).

While the SK concept is simple and easy to apply, the computation of the correct thresholds is more involved in the general case. To help with this, we include some example code written in the IDL language in Appendix~\ref{code}.  We note that the thresholds are typically just constants for data from a given instrumental setup.  For example, for the EOVSA case the $M=1792$, $N=1$ and $d=1$ values are fixed, hence the thresholds do not vary for the real-time \sk calculation.

\section*{Acknowledgments}

EOVSA was funded by a major construction grant from NSF under the American Recovery and Reinvestment Act. This work was supported by NSF grant AST-1312802, AFOSR grant FA9550-14-1-0336, and NASA grant NNX14AK66G to New Jersey Institute of Technology.

\newpage
\appendix{Code example for threshold computations}\label{code}
Here we provide an Interactive Data Language (IDL) example code that may be used to compute the generalized \sk thresholds with a predefined PFA based on equations provided by \citet{gsk}.  The program can be adapted to other languages provided replacements are available for IDL's incomplete gamma function (igamma) and NEWTON function, which is based on the routine newt described in section 9.7 of Numerical Recipes in C: The Art of Scientific Computing (Second Edition), published by Cambridge University Press.
\small{
\begin{verbatim}
;+
; NAME:
;       GSK
;
; PURPOSE:
;       This function computes the lower and upper detection thresholds
;       for the GSK estimator corresponding to a given false alarm (PFA) probability level
;
; CATEGORY:
;       Numerical Analysis.
;
; CALLING SEQUENCE:
;       Result = GSK(M,N=N,d=d,PFA=PFA)
;
; INPUTS:
;
; KEYWORD PARAMETERS:
;       N:      Number of on-board accumulations before S1 and S2 are computed.;
;               If not provided, N=1 is assumed.
;       d:      The shape parameter of the Gamma probability distribution corresponding to the PSD estimates.
;               The default value of d=1 corresponds to PSD estimates obtained by FFT means.
;               A value of d=0.5 corresponds to PSD estimates obtained by means of narrow
;               band filtering of a time domain signal.
;               Other vagues my be used to account for the particulars of the instrument
;       PFA:    False alarm probability level.
;               If not provided, the default PFA=0.0013499 (3-sigma normal PDF) is used.

; OUTPUTS:
;       This function returns the lower and upper thresholds in the array form [lower,upper]
;
; RESTRICTIONS:
;      Restricted to reasonably small values of M to avoid IGAMMA computation errors (M<1e9)
;
; PROCEDURE:
;
; EXAMPLES:
;
; REFERENCES:
;       Nita and Gary 2010, The Generalized Spectral Kurtosis Estimator,
;                           MNRAS, 406  (1):  L60-L64, doi: 10.1111/j.1745-3933.2010.00882.x
;
;
; MODIFICATION HISTORY:
;           Written April 2010 by Gelu M. Nita (gnita@njit.edu)
;
;-


FUNCTION upper_root, X
  common share, m1,m2,m3,m4,p
  return,abs((1-igamma(4*(m2^3)/(m3^2),(-(m3-2*m2^2)/m3+x[0])/(m3/2/m2)))-p)
END

FUNCTION lower_root, X
  common share, m1,m2,m3,m4,p
  return,abs(igamma(4*(m2^3)/(m3^2),(-(m3-2*m2^2)/m3+x[0])/(m3/2/m2))-p)
END



function GSK,M,N=N,d=d,PFA=PFA
  common share, m1,m2,m3,m4,p
  on_error,1

  ;Define default parameters
  if n_elements(N) eq 0 then N=1
  if n_elements(d) eq 0 then d=1
  if n_elements(M) eq 0 then M=6104
  if n_elements(PFA) eq 0 then PFA=0.0013499

  ;Force double precision
  p=double(pfa)
  M=double(M)
  Nd=double(N*d)
  PFA=double(PFA)

  ;Compute GSK moments according to equation 9 (used in threshold computation)
  m1=1d
  m2=(2*( M^2)* Nd *(1 + Nd))/((-1 + M) *(6 + 5* M* Nd + (M^2)*( Nd^2)))
  m3=(8*( M^3)* Nd* (1 + Nd)* (-2 + Nd* (-5 + M *(4 + Nd))))/(((-1 + M)^2)* $
    (2 + M* Nd) *(3 +M* Nd)* (4 + M* Nd)* (5 + M* Nd))
  m4=(12*( M^4)* Nd* (1 + Nd)* (24 + Nd *(48 + 84* Nd + $
    M *(-32 + Nd *(-245 - 93 *Nd + M* (125 + Nd* (68 + M + (3 + M)* Nd)))))))/$
    (((-1 + M)^3)* (2 + M* Nd)* (3 + M *Nd)* (4 + M* Nd) *(5 + M *Nd)* (6 + M* Nd)* (7 + M *Nd))

  ;Compute normalized moments entering Pearson criterion (not used below)
  beta1=(8* (2 + M *Nd)* (3 + M* Nd)* (-2 + Nd* (-5 + M* (4 + Nd)))^2)/$
    ((-1 + M)* Nd* (1 + Nd) *((4 + M *Nd)^2)* (5 + M *Nd)^2)
  beta2=(3* (2 + M *Nd) *(3 + M* Nd)* (24 + Nd* (48 + 84* Nd + M *(-32 + $
    Nd *(-245 - 93* Nd + M *(125 + Nd* (68 + M + (3 + M)* Nd)))))))/$
    ((-1 +M)* Nd *(1 + Nd)* (4 + M* Nd)* (5 + M *Nd)* (6 + M* Nd)* (7 + M *Nd))

  ;Compute Pearson criterion according to equation 11 (not used below)
  k=2*m2*(3*(m2^2)-m4)+3*(m3^2)

  ;Compute Type III parameters defined by equation 19
  delta=m1-2*(m2^2)/m3
  beta=4*(m2^3)/(m3^2)
  alpha=m3/(2*m2)

  ;Compute fourth moment error according to equation 21
  err4=abs((3* beta *(2 + beta)* (alpha^4)/m4-1)*100)

  ;Compute the thresholds according to equation 20

  upper = NEWTON([1], 'upper_root',/double,tolf=1e-8)
  lower = NEWTON([1], 'lower_root',/double,tolf=1e-8)

  return,[lower,upper]
end
\end{verbatim}
}
\bibliographystyle{ws-jai}

\bibliography{Nita_ads,ref,mssp}

\end{document}